\documentclass{ar-1col-S2O}
\usepackage[numbers]{natbib}
\usepackage{url}
\usepackage{amsmath}
\usepackage{amssymb}
\usepackage{epsfig}
\usepackage{graphicx}
\usepackage{color}
\usepackage[colorlinks,allcolors=blue]{hyperref}
\usepackage{physics}

\jname{Xxxx. Xxx. Xxx. Xxx.}
\jvol{AA}
\jyear{YYYY}
\doi{10.1146/((please add article doi))}

\begin{document}

% Page header
\markboth{Simenel et al.}{The Quest for Superheavy Nuclei: A Theoretical Perspective}

% Title
\title{The Quest for Superheavy Nuclei: A Theoretical Perspective}

%Authors, affiliations address.
\author{C. Simenel,$^1$ Lu Guo,$^2$ and K. Sekizawa$^{3,4,5}$
\affil{$^1$Department of Fundamental and Theoretical Physics, and Department of Nuclear Physics and Accelerator Applications, Research School of Physics, The Australian National University, Canberra ACT 2601, Australia; email: cedric.simenel@anu.edu.au}
\affil{$^2$School of Nuclear Science and Technology, University of Chinese Academy of Sciences, Beijing 100049, China; email: luguo@ucas.ac.cn}
%\affil{$^3$Institute of Theoretical Physics, Chinese Academy of Sciences, Beijing 100190, China}
\affil{$^3$Department of Physics, School of Science, Institute of Science Tokyo, Tokyo 152-8550, Japan}
\affil{$^4$Nuclear Physics Division, Center for Computational Sciences, University of Tsukuba, Ibaraki 305-8577, Japan}
\affil{$^5$RIKEN Nishina Center, Saitama 351-0198, Japan}
}

\%Abstract
\begin{abstract}
The synthesis of superheavy nuclei (SHN) lies at the forefront of nuclear physics, probing the limits of nuclear stability and the influence of shell effects.  
This review summarizes recent theoretical advances in understanding SHN formation and decay, emphasizing time-dependent density functional theory (TDDFT) and its extensions as microscopic tools to describe heavy-ion dynamics.  
Hybrid approaches combining TDDFT with coupled-channels, Langevin, and statistical models are discussed as means to connect microscopic predictions with experimental observables.  
The roles of fusion hindrance, quasi-fission, and multi-nucleon transfer are examined in terms of dissipation, shell structure, and deformation effects.  
Recent progress in computational power has enabled three-dimensional time-dependent mean-field (and beyond) simulations that include pairing and fluctuation dynamics.  
Perspectives are given on future developments toward a fully predictive description of superheavy element synthesis and the exploration of the upper limits of the nuclear landscape.
\end{abstract}

%Keywords, etc.
\maketitle

%Table of Contents
\tableofcontents

% Heading 1
\section{Introduction}

\begin{figure}
    \centering
    \includegraphics[width=0.9\linewidth]{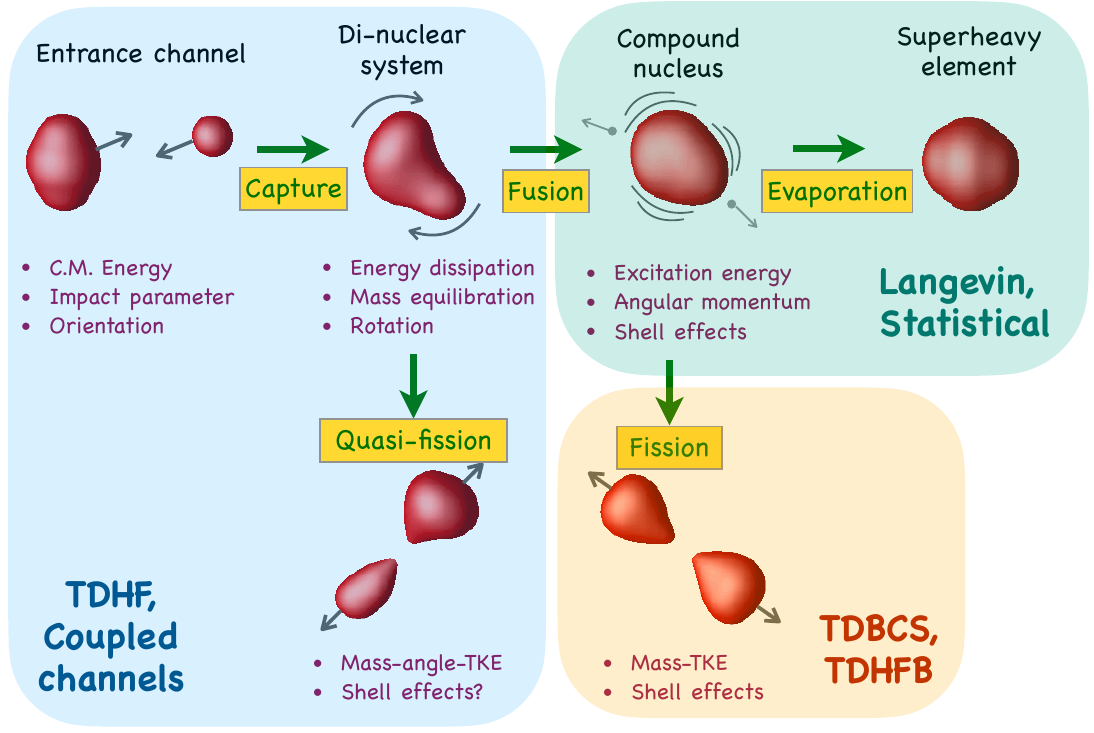}
    \caption{Various steps of superheavy element formation and competing reaction mechanisms. The dot points indicate the main quantities affecting each specific step. The theoretical approaches discussed in the manuscript are given (they do not provide an exhaustive list of models).}
    \label{fig:SHEflow}
\end{figure}

The quest for superheavy elements (SHE) and nuclei (SHN) has long attracted considerable interest within the nuclear physics community, from both theoretical and experimental perspectives.  
The complexity of the reaction mechanisms involved in SHN formation (See Fig.\ref{fig:SHEflow}) poses a major challenge for theorists aiming to provide quantitative predictions of production cross sections \cite{godbey2025}.  
\begin{marginnote}
\entry{SHE,SHN}
{Superheavy elements and nuclei have $Z\ge104$ protons.}
\end{marginnote}

The heaviest nuclei synthesized so far have been produced through fusion–evaporation reactions. While this mechanism is well understood and quantitatively described for medium-mass systems, it remains difficult to model for the heaviest nuclei.  
The formation of a cold residue involves successive stages—from the capture of the colliding nuclei to compound nucleus formation and statistical de-excitation via neutron evaporation.  
Quasi-fission strongly hinders compound nucleus formation, and fission largely dominates the decay process, making the accurate prediction of the rare survival channels particularly challenging.  
\begin{marginnote}
\entry{Compound nucleus}{In heavy-ion fusion, compound nuclei loose their projectile–target identity as high level density drives thermal equilibrium and statistical decay.}
\end{marginnote}

Multi-nucleon transfer reactions offer an alternative route to produce heavy and neutron-rich systems.  
However, they also present theoretical difficulties, including the determination of the properties of primary fragments (mass, charge, excitation energy, angular momentum, deformation) and the modeling of their decay, where neutron emission competes with fission.  

Time-dependent microscopic theories provide a powerful framework for investigating these processes, but their applicability is often limited by computational cost, restricting them to the mean-field level or modest extensions.  
Hybrid approaches that combine microscopic dynamics with phenomenological descriptions are therefore widely used: the early reaction stages leading to capture or primary fragment formation are treated microscopically, while subsequent de-excitation is handled by statistical models.  

This review summarizes the current theoretical developments in describing and predicting SHN synthesis and competing processes such as quasi-fission.  
Section~\ref{sec:model} introduces time-dependent density functional theory (TDDFT) and its extensions.  
Section~\ref{sec:entrance} examines entrance-channel dynamics and fusion competition, while Section~\ref{sec:CN} discusses compound nucleus formation and decay.  
Conclusions and perspectives are presented in Section~\ref{sec:conclusion}.
\begin{marginnote}
\entry{TDDFT}{In nuclear physics, the time-dependent density functional theory refers to microscopic dynamical approaches based on energy density functionals.}
\end{marginnote}

\section{Theoretical modelling \label{sec:model}}

\subsection{Time-dependent density functional theory}

TDDFT is used to simulate the dynamics of many-body systems.
In nuclear physics, it leads to the time-dependent Hartree-Fock (TDHF) equation and its extensions including pairing correlations.  
The Skyrme energy density functional (EDF) is often used to describe the interaction between nucleons. 
Indeed, it leads to time-dependent mean-field equations that are relatively easy to solve numerically using cartesian grids. 
\begin{marginnote}
\entry{EDF}{Energy density functionals express the energy as a functional of local densities and currents.}
\entry{TDHF}{The time-dependent Hartree-Fock theory provides a dynamical mean-field evolution of microscopic systems.}
\end{marginnote}

\begin{textbox}[h]
\section{Skyrme energy density functional with tensor terms}
The Skyrme effective interaction~\cite{skyrme1958} between nucleons is expressed as
\begin{eqnarray}
\begin{split}
\hat{v} (1,2) &=  t_0  \left( 1+x_0\, \hat{P}_{\sigma} \right) \delta + \frac{1}{2}  t_1  \left( 1+x_1\, \hat{P}_{\sigma} \right) 
\left(\hat{\bf k}'^2  \delta + \delta  \hat{\bf k}^2 \right) 
+ t_2  \left( 1+x_2\, \hat{P}_{\sigma} \right) 
\left(\hat{\bf k}' \cdot \delta  \hat{\bf k} \right) \\
&+ \frac{1}{6}  t_3 \left( 1+x_3 \hat{P}_{\sigma} \right) 
\rho^\alpha\! ({\bf \hat{R}})  \delta + i W_0 (\hat{\boldsymbol{\sigma}}_1+\hat{\boldsymbol{\sigma}}_2) \cdot \left(\hat{\bf k}' \times \delta  \hat{\bf k} \right) 
+\frac{1}{2}t_\mathrm{e}\Bigg\{\bigg[3\left(\hat{\boldsymbol{\sigma}}_1\cdot\hat{\bf k}'\right)\left(\hat{\boldsymbol{\sigma}}_2\cdot\hat{\bf k}'\right)\\
&-\left(\hat{\boldsymbol{\sigma}}_1\cdot\hat{\boldsymbol{\sigma}}_2\right)\hat{\bf k}'^2\bigg]\delta+\delta\bigg[3\left(\hat{\boldsymbol{\sigma}}_1\cdot\hat{\bf k}\right)\left(\hat{\boldsymbol{\sigma}}_2\cdot\hat{\bf k}\right)-\Big(\hat{\boldsymbol{\sigma}}_1\cdot\hat{\boldsymbol{\sigma}}_2\Big)\hat{\bf k}^2\bigg]\Bigg\}\\
&+t_\mathrm{o}\bigg[3\left(\hat{\boldsymbol{\sigma}}_1\cdot\hat{\bf k}'\right)\delta\left(\hat{\boldsymbol{\sigma}}_2\cdot\hat{\bf k}\right)-\Big(\hat{\boldsymbol{\sigma}}_1\cdot\hat{\boldsymbol{\sigma}}_2\Big)\hat{\bf k}' \delta \hat{\bf k}\bigg],
\label{eq:skyrme}
\end{split}
\end{eqnarray}
where $\delta = \delta\left(\hat{\bf r}_1-\hat{\bf r}_2\right)$, $\hat{\bf k} = \frac{1}{2i}\left(\nabla_1-\nabla_2\right)$, $\hat{\bf k}'$ is the complex conjugate of $\hat{\bf k}$ acting on the left, $\rho({\bf r})$ is the particle density,   $\bf \hat{R} = \left(\hat{\bf r}_1+\hat{\bf r}_2\right)/2$, and $\hat{\boldsymbol{\sigma}}_i$ is the usual vector of Pauli matrices.
%The operators $\ovsi = \ovsi(1)+ \ovsi(2)$, with
%$\ovsi(i) = \osi_x\!(i) \, \ve_x+ \osi_y\!(i) \, \ve_y + \osi_z\!(i) \, \ve_z$, 
%are expressed in terms of 
%The Pauli matrices $\osi_{x/y/z}(i)$ 
%act on the spin of the particle $i$ and
 $\hat{P}_\sigma = \left[1+  \hat{\boldsymbol{\sigma}}_1 \cdot \hat{\boldsymbol{\sigma}}_2 \right]/2$ exchanges the spins of the particles. 
The coupling constants $t_{0-3}$ terms describe the central force with an exchange character specified by $x_{0-3}$; the constant $W_{0}$ accounts for the short-range two-body spin-orbit force; and the $t_\textrm{e}$ and $t_\textrm{o}$ characterize the strengths of triplet-even and
triplet-odd tensor interactions, respectively. The full  Skyrme EDF is then written as
\begin{eqnarray}
\begin{split}
\mathcal{E} &= \mathcal{E}_0 + \sum_{\rm{t=0,1}} \left\{ 
C_{\rm{t}}^{\rm{s}} \mathbf{s}_{\rm{t}}^2 + C_{\rm{t}}^{\Delta{s}} \mathbf{s}_{\rm{t}} \cdot \Delta \mathbf{s}_{\rm{t}} 
+ C_{\rm{t}}^{\nabla s} \left(\nabla \cdot \mathbf{s}_{\rm{t}}\right)^2 
 + C_{\rm{t}}^{F} \left[ \mathbf{s}_{\rm{t}} \cdot \mathbf{F}_{\rm{t}} 
- \frac{1}{2} \left( \sum_{\mu=x}^{z} J_{\rm{t}, \mu\mu} \right)^{\!\!2}\right.\right.\\
&\left.\left.- \frac{1}{2} \sum_{\mu, \nu=x}^{z} J_{\rm{t}, \mu\nu} J_{\rm{t}, \nu \mu} \right] 
+ C_{\rm{t}}^{\rm{T}} \left( \mathbf{s}_{\rm{t}} \cdot \mathbf{T}_{\rm{t}} 
- \sum_{\mu,\nu=x}^{z} J_{\rm{t},\mu\nu} J_{\rm{t},\mu\nu} \right) \right\},
\end{split}
\label{EDFH}
\end{eqnarray}
where $\mathbf{s}, \mathbf{F},\mathbf{T}$ and $J$ denote the spin density, tensor-kinetic density, spin-kinetic density, and spin-current tensor density, respectively (see, e.g., \cite{perlinska2004,dai2014a} for definitions).
$\mathcal{E}_0$ denotes the standard functional without tensor force as used in most TDHF simulations. The remaining terms in Eq.~(\ref{EDFH}) arise from the tensor force.
\end{textbox}

Recently, all the time-even and time-odd tensor terms have been incorporated into modern TDHF simulations so that improved structure properties of both exotic and stable collision reactants, as well as their reaction dynamics are described in a unified microscopic way. The influence of tensor force has been extensively investigated in deep inelastic scattering~\cite{dai2014a}, fusion~\cite{stevenson2016,guo2018,guo2018b,stevenson2019,sun2022c,sun2022}, quasi-fission~\cite{li2022,li2024c}, and fission~\cite{huang2024b}.
It should be noted that the terms comprising the gradient of spin density which lead to  spin instability in both nuclear structure~\cite{Lesinski2007} and reaction dynamics~\cite{stevenson2016,guo2018} are usually neglected in numerical simulations.

\subsubsection{Time-Dependent Hartree-Fock}

The TDHF equation can be obtained by requesting the Dirac action
\begin{equation}
S[t_0,t_1;\Psi(t)]=\int_{t_0}^{t_1}dt \,\,\, \langle \Psi(t)| \left(i \frac{d}{dt}-\hat{H} \right)|\Psi(t)\rangle,
\label{eq:SDirac}
\end{equation}
to be stationary ($\delta S=0$) in the space of independent many-body states (Slater determinants) 
$|\phi\rangle =\prod_{i=1}^N\hat{a}^\dagger_i|-\rangle$,
where $\hat{a}^\dagger_i$ creates a particle in the state $|\varphi_i\rangle $ and $|-\rangle $ is the  vacuum.
This results in the TDHF equation (see, e.g., Appendix A in Ref.~\cite{simenel2025a} for details)
\begin{equation}
i\frac{\partial \rho}{\partial t}=\left[h[\rho],\rho\right],
\label{eq:tdhf}
\end{equation}
where the one-body density matrix elements are
$
\rho_{\alpha\beta}=\langle \phi| \hat{a}^\dagger_\beta\hat{a}_\alpha |\phi\rangle $.
The elements of the Hartree-Fock (HF) single-particle Hamiltonian $h[\rho]$ are defined  as
\begin{equation}
h_{\alpha\beta}=\frac{\delta \langle \phi| \hat{H}|\phi\rangle }{\delta \rho_{\beta\alpha}}.
\label{eq:hHF}
\end{equation}
In the DFT context, the expectation value of the Hamiltonian $ \langle \phi| \hat{H}|\phi\rangle$ is replaced by an EDF $\mathcal{E}$ (see Eq.~(\ref{EDFH}) for the Skyrme EDF), with the total energy $E[\rho]=\int d^3r\mathcal{E}$. 
The single-particle Hamiltonian $h[\rho]$ includes self-consistent mean-field potentials that are generated by all the particles. 
Each particle then evolves independently in this time-dependent potential.

\subsubsection{Extension of TDHF to describe pairing correlations} 

Although many works have been done in the context of SHE synthesis, the effects of pairing correlations, especially dynamical ones, have not been well explored to date. Dynamical effects of pairing correlations can be incorporated into time-dependent mean-field description based on the time-dependent Hartree-Fock-Bogoliubov (TDHFB) theory. 
\begin{marginnote}
\entry{Pairing correlations}{Neutrons (protons) form spin-singlet Cooper pairs, exhibiting properties of Fermionic superfluid (superconductor) in open-shell nuclei.}
\entry{TDHFB}{The time-dependent Hartree-Fock-Bogoliubov theory generalises TDHF equations to include dynamical pairing correlations.}
\end{marginnote}

\begin{textbox}[t]
\section{TDHFB formalism}
In TDHFB and its static version (HFB), the trial wave function of the variation is extended to the vacuum state of quasiparticles $|\text{HFB}\bigr>$ defined by 
\begin{equation}
\hat{\beta}_\mu\bigl|\text{HFB}\bigr> = 0,
\end{equation}
where $\hat{\beta}_\mu$ (and $\hat{\beta}_\mu^\dagger$) represent the quasiparticle annihilation (creation) operators,
\begin{eqnarray}
\hat{\beta}_\mu &=& \sum_k\Bigl(U_{k\mu}^*\hat{a}_k+V_{k\mu}^*\hat{a}_k^\dagger\Bigr),\\
\hat{\beta}_\mu^\dagger &=& \sum_k\Bigl(U_{k\mu}\hat{a}_k+V_{k\mu}\hat{a}_k^\dagger\Bigr).
\end{eqnarray}
The variational principle leads to the TDHFB equation,
\begin{equation}
i\frac{\partial\mathcal{R}}{\partial t} = \bigl[\mathcal{H}[\rho,\kappa],\mathcal{R}\bigr],
\end{equation}
where $\mathcal{H}$ and $\mathcal{R}$ are the HFB matrix and the generalized density matrix, respectively, which are given by
\begin{eqnarray}
\mathcal{H} = \begin{pmatrix}h&\Delta\\-\Delta^*&-h^*\end{pmatrix},\quad
\mathcal{R} = \begin{pmatrix}\rho&\kappa\\-\kappa^*&I-\rho^*\end{pmatrix}.
\end{eqnarray}
Here, $\kappa$ is the anomalous density matrix whose matrix elements are given by $$\kappa_{\alpha\beta}\equiv\bigl<\text{HFB}\big|\hat{a}_\beta\hat{a}_\alpha\big|\text{HFB}\bigr>=(V^*U^\text{T})_{\alpha\beta}.$$ Note that the anomalous density can take non-zero values because the quasiparticle vacuum is a complex mixture of creation and annihilation operators.
\end{textbox}

By solving the TDHFB equation, one can include the effects of pairing correlations in mean-field dynamics. In practical applications, however, there is a problem of computational costs. It is because of the fact that to solve the TDHFB equation one has to evolve a huge number of quasiparticle states, which could be 1000 times or more larger than the number of nucleons in the system. It has thus become possible only recently to directly solve the TDHFB equation in three-dimensional coordinate space using top-tier supercomputers. Recent applications indicate new dynamic effects of pairing, for instance, soliton-like excitations in collisions of superfluid nuclei \cite{magierski2017} and the  emergence of pairing correlations in fusing $^{90}$Zr+$^{90}$Zr system \cite{magierski2022}. Further applications of TDHFB for SHE synthesis is desired, which may potentially uncover unexpected effects of pairing correlations in their synthesis.

On the other hand, an approximate treatment of pairing correlations has been widely used in the nuclear physics community which is called TDHF+BCS, TDBCS, or canonical-basis TDHFB approaches. It has been shown \cite{ebata2010} that TDHF+BCS can be derived from TDHFB by working with the canonical basis that diagonalizes the one-body density matrix and by assuming that the pairing tensor takes non-zero values for the BCS-type components only, i.e.\ $\Delta_{k\bar{k}}=-\Delta_{\bar{k}k}$, for all times, where $k$ and $\bar{k}$ are time-reversal partners. It was shown in Ref.~\cite{ebata2010} that  the linear-response limit of TDHFB [i.e.\ quasiparticle random phase approximation (QRPA)] and TDHF+BCS provide identical results. 
\begin{marginnote}
\entry{TDHF+BCS}{The Bardeen-Cooper-Schrieffer approximation is often used to simplify TDHFB equations.}
\end{marginnote}

The TDHF+BCS approach has been extensively used for describing fusion and transfer processes in low-energy heavy-ion reactions \cite{scamps2013a} as well as dynamics of induced fission processes \cite{scamps2015a,scamps2018,scamps2019}. We should note, however, that TDHF+BCS is an approximate formalism (e.g., it cannot describe spatial modulations of the pairing gap, and it violates the continuity equation \cite{scamps2012}). Detailed analyses are required to assess to which extent the simplified TDHF+BCS description is justified, if one avoids employing full TDHFB.

\subsubsection{Density constrained methods \label{sec:DC}}

The internuclear potential can  be calculated microscopically
with the (TD)DFT approach by applying
frozen Hartree-Fock (FHF)~\cite{simenel2008,guo2012,simenel2013b,bourgin2016,vophuoc2016},
density-constrained FHF (DCFHF)~\cite{simenel2017,umar2021},
DC-TDHF~\cite{umar2006b}, or
dissipative-dynamics TDHF~\cite{washiyama2008} approaches.
To preserve the consistency with microscopic calculations, it is necessary to compute the potential from the same EDF used in the static HF calculations of nuclear ground-states.
The obtained potential can be used to calculate penetration probabilities with the incoming wave boundary condition method
\cite{hagino1999}.
In this section, we will introduce the FHF, DCFHF, and DC-TDHF methods.
\begin{marginnote}
\entry{DC-TDHF}{The density constrained TDHF method provides nucleus-nucleus potentials accounting for dynamical effects and the Pauli exclusion principle.}
\entry{DCFHF}{The density constrained frozen Hartree-Fock method neglects dynamical rearrangement of the density but still accounts for Pauli exclusion principle.}
\entry{FHF}{The frozen Hartree-Fock method neglects dynamics and the Pauli exclusion principle between nucleons of different collision partners.}
\end{marginnote}

Since the TDHF theory describes the collective motion of fusion dynamics in terms of semi-classical trajectories, the sub-barrier quantum
tunneling of the many-body wave function cannot be included. Consequently, direct TDHF calculations cannot be used to describe sub-barrier fusion.
At present, all sub-barrier fusion calculations assume that there exists an ion-ion potential which depends on the internuclear distance.

\begin{textbox}[t]
\section{Density-constrained TDHF}
The microscopic DC-TDHF approach~\cite{umar2006b} is employed to extract the nucleus-nucleus potential from the TDHF time evolution of the dinuclear system.
In this approach, at certain time during the evolution, the instantaneous TDHF density is used to perform a static HF energy minimization
\begin{equation}
\delta \langle \Psi_{\rm DC}|\hat{H}-\int d^3r \lambda(\textbf{r})\rho(\textbf{r})|\Psi_{\rm DC}\rangle=0,
\end{equation}
by constraining the proton and neutron densities to be equal to the instantaneous TDHF densities. Since the local density is constrained,
all moments are simultaneously constrained. DC-TDHF calculations give the adiabatic reference state for a given TDHF state, which is the Slater
determinant with the lowest energy for a given density.
The minimized energy
\begin{equation}
E_{\rm {DC}}(\textbf R)=\langle \Psi_{\rm DC}|H|\Psi_{\rm DC}\rangle
\end{equation}
is the density-constrained energy.
Since this density-constrained potential still contains the binding energies of individual nuclei which should be subtracted out,
the heavy-ion interaction potential is deduced as
\begin{equation}
V(\textbf R)=E_{\rm {DC}}(\textbf R)-E_{\rm {A}_1}-E_{\rm {A}_2},
\label{VB}
\end{equation}
where $E_{\rm {A}_1}$ and $E_{\rm {A}_2}$ are the binding energies of the two individual nuclei.
One should note that this procedure does not affect the TDHF time evolution and contains no free parameters or normalization.
\end{textbox}

The DC-TDHF technique has been introduced to compute the nucleus-nucleus potential in a dynamical microscopic way.
All of the dynamical effects included in TDHF are then directly incorporated in the potential. However, it is sometimes desirable to look for a different approach
to produce a bare potential which does not include any dynamical contribution. This is useful, for instance, if one wants to couple TDHF with coupled-channel calculations in order to avoid double counting the effect of the dynamics. A bare potential also allows to disentangle the static and dynamical effects, e.g., 
of the tensor force. The bare nucleus-nucleus potential is defined as the interaction potential between the nuclei in their ground states.
%This is possible using the frozen Hartree-Fock (FHF) technique~\cite{simenel2013b}, assuming that the densities of the target and projectile remain constant and equal to their respective ground state densities. The potential can then be expressed as
%\begin{align}
%\label{eq:FD}
%V_\mathrm{FD}(\textbf R)=E[\rho_\mathrm{1}+\rho_\mathrm{2}](\textbf R)-E[\rho_\mathrm{1}]-E[\rho_\mathrm{2}],
%\end{align}
%where $\rho_\mathrm{1}$ and $\rho_\mathrm{2}$ are HF ground state densities of the fragments, and $E[\rho_\mathrm{1}+\rho_\mathrm{2}]$ is obtained from the Skyrme EDF. In the FHF approach, the Pauli principle between pairs of nucleons belonging to different collision partners has been neglected. When the overlap between the density distributions is small, the barrier is almost unaffected by the inclusion of the Pauli principle. However, at larger overlaps of the densities where the Pauli principle is expected to play a more important role, the FHF approximation may not properly account for the potential, particularly the inner part~\cite{simenel2017}.

Comparing with the internuclear potentials from DC-TDHF,
those from the DCFHF method do not include any dynamic factors
and the contribution from Pauli exclusion principle is still included.
The Pauli exclusion principle is included by allowing the single-particle states
to reorganize to attain minimum energy 
in the static HF calculations with the density constraint and to
be properly antisymmetrized, as the many-body state is a
Slater determinant of all the occupied single-particle wave
functions.

\begin{textbox}[h]
\section{DCFHF and FHF}
The FHF calculations are preformed with constraining 
the total proton $p$ and neutron $n$ densities to be
the same as those at the ground state,
\begin{equation}
	\delta \left \langle
	\hat{H}-\int \text{d}^3r \sum_{q=p,n} \lambda_q(\textbf{r})
	\left[\rho_{\rm{P},q} ({\bf r}) +
	\rho_{\rm{T},q} (\bf r- \bf R)
	\right]\right\rangle=0,
\end{equation}
where $\rho_{\rm{P}}$ and $\rho_{\rm{T}}$ are the densities of
the projectile and target in their ground states, respectively. This variational procedure results in a unique
Slater determinant $\Phi({\bf R})$. Similar to the case of DC-TDHF,
the internuclear potential from DCFHF is given by
\begin{equation}
	V_\mathrm{DCFHF}({\bf R})=\langle \Phi({\bf R})|\hat{H}|\Phi({\bf R})\rangle -E_{\rm {A}_1}-E_{\rm {A}_2}.
	\label{VB1}
\end{equation}

The FHF potential is expressed as
\begin{equation}
\label{eq:FD}
V_\mathrm{FHF}({\bf R})=E[\rho_\mathrm{P}+\rho_\mathrm{T}]({\bf R}) -E_{\rm {A}_1}-E_{\rm {A}_2},
%-E[\rho_\mathrm{P}]-E[\rho_\mathrm{T}].
\end{equation}
where $E[\rho_\text{P}+\rho_\text{T}]({\bf R})$ denotes the energy of the system where the ground-state densities of projectile and target nuclei, $\rho_\text{P}$ and $\rho_\text{T}$, are placed with the relative vector ${\bf R}$.
\end{textbox}

%The DC-TDHF method  has been introduced to compute the nucleus-nucleus potential in a dynamical microscopic way. All of the dynamical effects included in TDHF and Pauli exclusion principle are then directly incorporated. In DCFHF, the dynamic effects are neglected while the Pauli exclusion principle is still included.
There is also a potential
which includes neither dynamical effects nor the contributions of Pauli exclusion principle.
This nucleus-nucleus potential is defined as the potential between the nuclei in their ground states.
This is achieved with the FHF technique \cite{simenel2013b},
assuming that the densities of the target and
projectile are unchanged and equal to their ground state densities.
In particular, FHF potentials are useful to isolate the contribution from Pauli repulsion by comparison with DCFHF potentials \cite{simenel2017,umar2021}. 

When the internuclear distance $R$ is large, i.e., the overlap between the densities of the projectile and target is small, the Pauli principle
almost has no influence on the internuclear potential.
However, when two nuclei are close to each other and the density overlaps are large, the Pauli principle is expected to play an important role.
This is particularly true for reactions used to form superheavy nuclei as the nuclear wave-functions of the collision partners need a significant spatial overlap to reach the fusion barrier due to the large Coulomb repulsion between the reactants.
Therefore the FHF approximation cannot properly
describe the inner part of the potential (cf.\ Ref.~\cite{simenel2017}) in heavy-ion collisions relevant to SHE synthesis.

\subsubsection{Numerical aspects}

%\begin{textbox}[h]
%\section{3D TDDFT solvers}
Several solvers have been developed to simulate nuclear dynamics within TDDFT in three dimensions (see Tab.~\ref{tab:codes}).
\begin{table}
\caption{Three-dimensional TDDFT solvers used in heavy-ion collisions studies.}
\label{tab:codes}
\begin{tabular}{lccccc}
Solver 									& EDF			& TDRPA 						& Dynamical pairing 				& Tensor\\
\hline
\textsc{tdhf3d}\footnote{\textsc{tdhf3d} assumes a plane of symmetry.} 
			 \cite{kim1997}					& Skyrme				& \cite{simenel2011}				& TDHF+BCS \cite{scamps2013a} 	& \\
\textsc{sky3d}  \cite{abhishek2024}				& Skyrme				& \cite{gao2025}				& 							&  \cite{dai2014a,stevenson2016}\\
\textsc{hit3d}  \cite{shi2024}					& Skyrme				&							& 							& \cite{shi2024}\\
Vanderbilt code \cite{umar2005a}				& Skyrme				& \cite{godbey2020b}			& 							& \\
Tsukuba code \cite{nakatsukasa2005,sekizawa2013}& Skyrme				& \cite{williams2018}				& 							& \\
\textsc{lise}  \cite{jin2021}						& Skyrme				&							& TDHFB						& \\
 Beijing code \cite{ren2020}					& Covariant			&							& TDHF+BCS \cite{ren2022}		& \\
 Tsukuba (Gogny) code \cite{hashimoto2016}		& Gogny				&							& TDHFB						& \\
\end{tabular}
\end{table}

The collision of two nuclei is typically simulated on a 3D Cartesian grid using time-iterative methods. First, static calculations determine each nucleus’s ground state. The nuclei are then placed in a larger box without overlap, forming an independent (quasi)particle state. A Galilean boost sets their initial motion, assuming a prior Rutherford trajectory. The mean-field equations are then solved iteratively, and one-body observables are tracked over time. 

While most EDFs are fitted with center-of-mass (c.m.) corrections, these are not included in mean-field simulations of heavy-ion collisions as they introduce a spurious dependence on the nucleon numbers of each fragment. 
To treat structure and dynamics on the same footing, Skyrme EDF such as SLy4$d$ \cite{kim1997} and UNEDF1 \cite{kortelainen2012} should be used as they have not been fitted with centre of mass corrections. Although SLy4d has been more popular for TDHF reaction studies, UNEDF1 has also been used in fusion studies \cite{vophuoc2016,godbey2022}. Unlike SLy4d, it allows for  uncertainty quantification \cite{godbey2022}. One should keep in mind, however, that fitting nuclear ground-state properties without the c.m.\ corrections causes spurious surface properties \cite{bender2000cmcorrection,dacosta2024,kafker2025}, which may affect description of shape evolution dynamics in (quasi)fission and fusion, including SHE synthesis.

The mean-field equations are usually solved in a cartesian grid with mesh spacing $\Delta x=0.6$--$1.0$~fm. As time-reversal symmetry cannot be assumed, all single-particle wave-functions with a non-negligible occupation number are evolved in time.
Due to the self-consistency of the mean-field, small time-steps of the order of $\Delta t\approx 10^{-24}$~s are used. A typical evolution for the description of a fusion reaction can require few thousands time iterations. 

While TDHF+BCS codes are simple extensions of existing TDHF ones including a time evolution of the occupation numbers \cite{scamps2013a}, TDHFB solvers have been developed on their own. 
Nevertheless, similar algorithms are used for the time evolution of the (quasi)particle wave-functions. 
However, TDHFB codes require significantly more computational power~\cite{jin2021}.

\subsection{Coupling TDHF to capture, compound nucleus formation and decay models}

Due to its limitations, in particular the lack of quantum tunneling and two-body mechanisms, TDHF is not able to describe the complete evolution from two colliding partners to the formation of a fully equilibrated compound nucleus and its subsequent decay. To overcome these limitations, TDHF is often coupled to complementary models. One approach is to use TDHF to determine the entry point (where initial kinetic energy is fully dissipated) on a (multi-dimensional) potential energy surface (PES), and then determine the subsequent evolution with, e.g., the Langevin equation \cite{sekizawa2019b}. In addition, although TDHF accounts for effects of couplings between intrinsic degrees of freedom and relative motion {\it on average}, it does not allow for distribution of potential barriers induced by such couplings. TDHF is then sometimes completed by coupled channel calculations \cite{simenel2013b,sun2022b}. Here, we describe briefly various theoretical ingredients entering such calculations. 
\begin{marginnote}
\entry{PES}{Potential energy surfaces represent the evolution of the potential energy with nuclear shape such as elongation and asymmetry.}
\end{marginnote}

\subsubsection{Potential energy surface}

PES can be determined with the same EDF as in TDHF calculations. However, they are often computed with additional approximations, namely axial symmetry and zero angular momentum and temperature. Using TDHF trajectories, which usually break these approximations, to determine the entry point of a colliding system on the PES, thus needs to be done with caution. Nevertheless, PES has been successfully used to interpret TDHF simulations of quasi-fission, thus demonstrating that PES topography remains relevant to describe, at least qualitatively, the heavy-ion collision dynamics at near barrier energies \cite{simenel2021,lee2024b,mcglynn2023,simenel2025c}. 

PES can be obtained by solving constrained static mean-field equations. It provides a representation of the potential energy of the system for given shapes characterised by, e.g., multipole moments. In particular, the quadrupole moment $Q_{20}$ is a proxy for the elongation of the system while the octupole moment $Q_{30}$ quantifies its asymmetry. Higher multipole moments are usually left unconstrained. PES is a standard concept entering theoretical fission studies (see, e.g., \cite{bender2020}).

\begin{figure}
    \centering
    \includegraphics[width=0.7\linewidth]{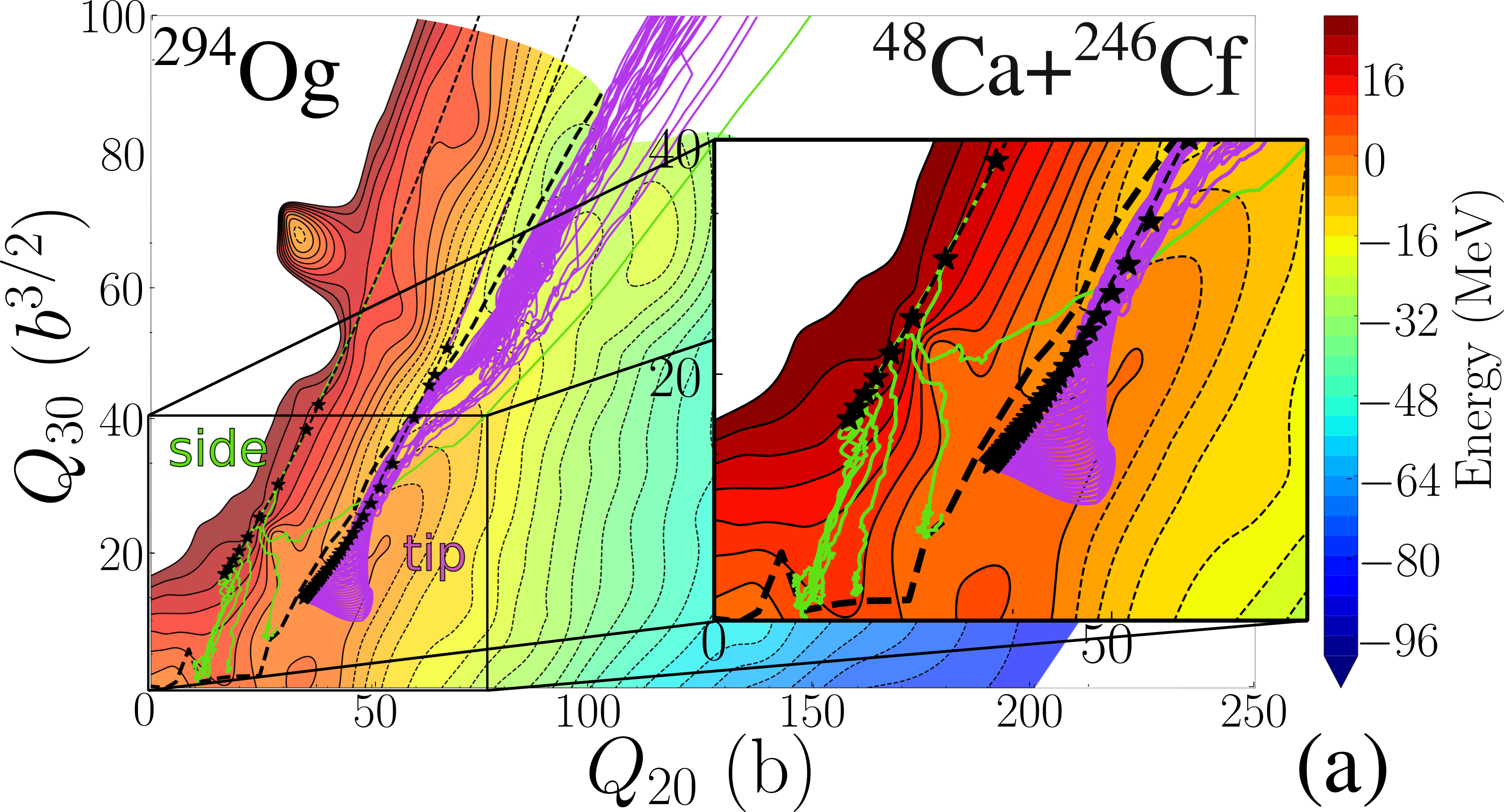}
    \caption{Overlay of TDHF trajectories of $^{48}$Ca$+^{246}$Cf on the PES of the compound nucleus $^{294}$Og. 
    The PES represents the potential energy (see color bar) of  $^{294}$Og as function of quadrupole ($Q_{20}$) and octupole ($Q_{30}$) moments quantifying elongation and asymmetry, respectively. 
     Entry points (defined here as neck density reaching 0.08~fm$^{-3}$) are represented by stars. 
   The short-dashed lines connected to the entry points correspond to the incident kinematic trajectories.  
    Solid green lines (collisions with the side of $^{246}$Cf) and solid purple lines (collisions with the tip) show the trajectories from the entry point. 
	The long-dashed line represents the one-dimensional adiabatic fission path. The inset is a zoom of the compact region of the PES. From Ref.~\cite{mcglynn2023}.}
    \label{fig:294Og}
\end{figure}

Figure \ref{fig:294Og} shows an example of PES for the superheavy $^{294}$Og nucleus \cite{mcglynn2023} obtained with the \textsc{skyax} code \cite{reinhard2021} and SLy4d Skyrme EDF \cite{kim1997}. The adiabatic one-dimensional fission path represented by the dashed line follows the so-called super-asymmetric fission valley discussed in Sec.~\ref{sec:fission}. The use of such PES to interpret TDHF trajectories in the context of fusion hindrance and quasi-fission is discussed in Sec.~\ref{sec:entrance}.

\subsubsection{Coupled-channels model}

Coupled-channels calculations have been combined with TDHF simulations in several applications to fusion \cite{simenel2013a,simenel2013b,umar2014a,bourgin2016,sun2022b}. Indeed, while dynamical effects such as the coupling between relative motion and internal excitations is accounted for in TDHF at the mean-field level, i.e., affecting the average fusion barrier, it does not account for the full distribution of fusion barriers that are better described within the coupled-channels formalism. 
\begin{marginnote}
\entry{Coupled-channels model}{Quantum method to account for the effects of nuclear excitations on nucleus-nucleus potentials coherently.}
\end{marginnote}

\begin{textbox}[h]
\section{Coupled-channel approach}
In the coupled-channels framework, the relative motion between the interacting nuclei is governed by an effective potential that combines both the nuclear and Coulomb interactions. 
A commonly adopted parametrization for the nuclear part of the nucleus–nucleus potential is given by the Woods–Saxon form
\begin{equation}
V_{WS}(D) = \frac{-V_0}{1 + \exp\left[\dfrac{D - r_0(A_1^{1/3} + A_2^{1/3})}{a}\right]},
\label{eq:WS}
\end{equation}
where $V_0$ denotes the potential depth and $a$ its diffuseness parameter.

Excitations of vibrational modes lead to fluctuations in the distance between the nuclear surfaces. 
These effects can be incorporated by replacing the mean radii of the colliding partners, $R_{0_i}=r_0A_i^{1/3}$, in Eq.~(\ref{eq:WS}) with the operator $\hat{R}_i(\theta,\phi)$ that represents the distance to the nuclear surface of nucleus $i$. 
This surface can, for instance, be defined as the isodensity contour corresponding to $\rho_0/2$, with $\rho_0 = 0.16$~fm$^{-3}$ being the nuclear saturation density. 
The surface operator can then be expressed as \cite{hagino2012}
\begin{equation}
\hat{R}(\theta,\phi) \simeq R_0 + \sum_{\lambda \ge 2} \sum_\nu 
\frac{R_0 \, \beta_{\lambda}^{(\nu)}}{\sqrt{2\lambda + 1}}
\left( \hat{a}^{\dagger(\nu)}_{\lambda\mu} + (-)^{\mu}\hat{a}^{(\nu)}_{\lambda\mu} \right)
Y_{\lambda\mu}^*(\theta,\phi),
\label{eq:oR}
\end{equation}
where $\beta_{\lambda}^{(\nu)}$ is the deformation parameter associated with the phonon state $|\nu\rangle$. 
The operators $\hat{a}^{\dagger(\nu)}_{\lambda\mu}$ and $\hat{a}^{(\nu)}_{\lambda\mu}$ create and annihilate, respectively, a phonon $|\nu\rangle$ carrying angular momentum $\lambda$ and projection $\mu$. 
Within the isocentrifugal approximation, the spherical harmonics in Eq.~(\ref{eq:oR}) reduce to their $\mu=0$ component, leading to a simplified description of the couplings \cite{hagino2012}. 
The nuclear potential including all coupling orders is then obtained by replacing the coordinate in the Woods–Saxon potential as
\begin{equation}
\hat{V} \equiv V_{WS}\!\left(\hat{D} - \sum_{i=1}^{2} R_{0_i}
\sum_{\nu, \lambda \ge 2} 
\frac{\beta_{\lambda_i}^{(\nu)}}{\sqrt{4\pi}}
\left( \hat{a}^{\dagger(\nu)}_{\lambda0}(i) + \hat{a}^{(\nu)}_{\lambda0}(i) \right)
\right).
\nonumber
\end{equation}
\end{textbox}

In principle, the coupled-channel formalism allows one to account for the influence of both low-lying collective vibrations and high-energy giant resonances. 
Low-lying collective excitations tend to fragment the single potential barrier, giving rise to a distribution of barriers \cite{dasgupta1998}. 
In contrast, couplings to high-lying collective modes, such as giant resonances, primarily cause a global shift of the barrier distribution without significantly altering its shape \cite{hagino1997}. 
A similar effect occurs in light systems, where the small product $Z_1Z_2$ yields weak coupling strengths. 
%For example, the coupling to the $3^-_1$ state of $^{40}$Ca in reactions with light partners, or of $^{16}$O with any target nucleus, mainly results in an adiabatic renormalization of the static potential without modifying the structure of the barrier distribution \cite{das98}.

In standard coupled-channels calculations—such as those performed with the code \textsc{ccfull} \cite{hagino1999}—the collective model, employing deformation parameters and excitation energies extracted from experiment, is used to reproduce the {\it shape} of the empirical barrier distribution. 
Meanwhile, the parameters of the nucleus–nucleus potential are fine-tuned to reproduce its {\it centroid}. 
In applications to SHN formation, however, the centroid of the barrier distribution is not always known and one has then to rely on microscopically derived potentials as discussed in Sec.~\ref{sec:DC}.

\subsubsection{Langevin method} 

Although TDHF has been successful in describing a variety of dynamic phenomena in a microscopic way, there are many limitations inherent to the mean-field approximation. One of such limitations is that it cannot describe the compound nucleus formation process, because nucleon-nucleon collisions are absent in TDHF and thermalization of a composite system cannot be fully accounted for.
\begin{marginnote}
\entry{Langevin equation}{A stochastic equation of motion that incorporates damping and thermal noise to model dissipative dynamics.}
\end{marginnote}

It is therefore necessary to combine TDHF with some other model that can describe the compound nucleus formation process after capture. As the first step towards this direction, a combination of TDHF and the so-called fusion-by-diffusion (FbD) model has been proposed \cite{sun2022b,sun2023}. In the FbD model, the compound nucleus formation process is described as an up-hill diffusion over the one-dimensional inner barrier, assuming that the system evolves along a valley in the multi-dimensional PES, where all degrees of freedom are equilibrated in a much shorter timescale than that of the up-hill diffusion process. In the FbD model, the shape of the inner barrier is expressed as an inverted parabola, which allows one to analytically solve the diffusion equation, providing an analytical expression of the compound nucleus formation probability, $P_\text{CN}$. The only input of the FbD model is the so-called ``injection point'' (or entry point) which characterizes the initial elongation where the up-hill diffusion process starts. In Ref.~\cite{sekizawa2019b}, hot fusion reactions that can lead to formation of the element 120 have been analyzed, where the initial stage of the reactions are described in TDHF, determining the injection point from the distance of closest approach after collision. The FbD model was then used to calculate $P_\text{CN}$, and a statistical model was separately used to evaluate the survival probability.
\begin{marginnote}
\entry{FbD}{A phenomenological model where heavy-ion fusion occurs by diffusion over an inner fission barrier after nuclear contact.}
\end{marginnote}

%This kind of combinations of TDHF with other models is promising to extend applications of microscopic approaches for superheavy element synthesis. %So far, a simple FbD model has been used in the literature, but other theoretical models could be used, e.g., the random walk model \cite{randomwalk} or the multidimensional Langevin model \cite{langevin}. 
%By determining input quantities in phenomenological models through microscopic TDHF calculations, the predictive power of the formers will be enhanced.

\subsubsection{Statistical decay} 

Using the TDHF approach one can calculate reaction dynamics in a relatively short timescale up to around tens of zeptoseonds. The upper limit is set by not only computational costs required for a longer time evolution but also by integrated numerical inaccuracies over  time. Together with the absence of thermalization in TDHF, one has to rely on a statistical model to evaluate effects of secondary disintegration processes such as neutron evaporation and fission. Indeed, in the above mentioned studies that employ the combination of TDHF with the FbD model, a statistical model was used to evaluate the survival probability of the compound nucleus formed.

The same is true also for the case of binary reaction products. After deep-inelastic collisions that include multi-nucleon transfer and quasi-fission processes, reaction products can be highly excited. Often, experimental cross sections include effects of secondary disintegration processes, and one needs to employ a statistical model to make a direct comparison with experimental data. In Ref.~\cite{sekizawa2017}, such a combination was achieved, where production cross sections of primary fragments are calculated by employing the particle-number projection method \cite{simenel2010} and the effects of secondary disintegration processes are included by employing the statistical model \texttt{GEMINI++}. In the statistical model, excitation energy and angular momentum, as well as neutron and proton numbers of the decaying nucleus are required. Those quantities can be evaluated within TDHF either by taking expectation values in a particle-number projected state \cite{sekizawa2014} or by simply taking average values in TDHF. %Later, such a combination of TDHF with \texttt{GEMINI++} has been used in literature \cite{sekizawa2017}.

\section{Entrance channel mechanisms in competition with fusion \label{sec:entrance}} 

\subsection{Fusion hindrance}

Fusion reactions producing heavy and superheavy nuclei require an extra-push energy over the fusion barrier. This extra-push energy is necessary to overcome dissipation and the strong Pauli repulsion. To investigate extra-push dynamics in a fully microscopic scenario, one may compare the lowest bombarding energy required to achieve fusion for a central TDHF collision with the fusion barrier in the nucleus-nucleus interaction potential. The potential may be approximately obtained with the FHF method using the same EDF as TDHF (see Eq.~(\ref{eq:FD})).
The difference between the TDHF fusion energy threshold at zero impact parameter and the FHF barrier height provides an estimate for the additional extra-push energy required for fusion. 
\begin{marginnote}
\entry{Extra-push}{Additional energy above the Coulomb barrier required to achieve fusion in heavy nuclear systems.}
\end{marginnote}

Fusion barriers and low-energy fusion thresholds have been systematically calculated 
for reactions involving combinations of spherical nuclei
$^{16}\rm O$, $^{40}\rm Ca$, $^{48}\rm Ca$, $^{48}$Ti, $^{90,96}\rm Zr$, $^{100,124,132}\rm Sn$, $^{136}$Xe, and $^{208}\rm Pb$ \cite{guo2012,simenel2025a}. In addition, 
 $^{48}\rm Ca+^{238}\rm U$, $^{96}\rm Zr+^{132}\rm Zr$, and $^{70}\rm Zn+^{208}\rm Pb$,  leading to the synthesis of superheavy elements,
have been investigated  \cite{guo2012}. 
%The  difference between TDHF low-energy fusion threshold $E_{\rm lf}^{\rm TDHF}$ 
%and the fusion barrier obtained from the FHF method $V_{\rm B}^{\rm FD}$ are reported in Fig.~\ref{fig:45}. 
%The horizontal axis is 
The resulting behavior of the extra-push varies with the effective fissility 
$(Z^2/A)_{\rm eff} \equiv 4Z_1Z_2/A_1^{1/3}A_2^{1/3}(A_1^{1/3}+A_2^{1/3})$.
%The quantity $E_{\rm lf}^{\rm TDHF}-V_{\rm B}^{\rm FD}$ is shown in blue line. 
For light systems with effective fissility smaller than 33, %this quantity is negative, meaning that 
TDHF fusion thresholds are smaller than FHF fusion barriers.
This is interpreted as an effect of couplings between relative motion and internal collective degrees of freedom that, in average, lower the TDHF fusion threshold \cite{simenel2013a,simenel2013b,umar2014a}. 
%\begin{figure}
%\resizebox{1.0\columnwidth}{!}{\includegraphics{fig5}}
%\caption{Energy difference between TDHF low-energy fusion threshold and fusion barrier with FHF method (blue line) compared with the results from the extra-push formula in Swiatecki phenomenological model (purple line). From Ref.~\cite{guo2012}.}
%\label{fig:45} 
%\end{figure}
%Figure \ref{fig:45} also shows that 
However, an extra-push energy is  predicted from the TDHF calculations for $(Z^2/A)_\text{eff}>33$.
 This prediction on the onset of extra push is consistent with Swiatecki's macroscopic model~\cite{swiatecki1982}. %However, unlike Swiatecki's predictions,  the extra-push energy in TDHF does not monotonically  increase as a function of effective fissility, as seen from the staggering of blue line. 
%Since the collision dynamics is affected by many factors, e.g., nuclear structure and dynamical effects in the time evolution, the extra-push
%energy will not simply depend on the mass or charge combination. 
However, the magnitude of the extra-push energy is  significantly smaller in the microscopic approach than Swiatecki's predictions.
In general, TDHF prediction of fusion thresholds is in excellent agreement with centroids of experimental barrier distributions \cite{simenel2008,guo2012}, even for systems requiring an extra-push energy to fuse. 

The need for an extra-push is due to the fact that the entry point in the PES for heavy systems often lies beyond the conditional saddle point. We see in Fig.~\ref{fig:294Og} that for collisions of $^{48}$Ca with the tip of $^{246}$Cf, the entry points are located at larger elongations  than the saddle point marking the start of the superasymmetric fission valley at $Q_{20}\simeq30$~b and $Q_{30}\simeq10$~b$^{2/3}$. Collisions with the side of $^{246}$Cf, however, have entry points that are found at smaller values of $Q_{20}$ leading to TDHF trajectories trapped inside the conditional saddle point, i.e., fusion (see inset of Fig.~\ref{fig:294Og}). A proper account of deformation and orientation effects on the dynamics is then crucial for evaluation of extra push energies and quasi-fission probabilities \cite{hinde2018}.

\subsection{Multi-nucleon transfer} 

In applications of TDHF for heavy-ion reactions, it was customary to calculate average reaction outcomes such as total kinetic energy loss (TKEL) to quantify dissipation and study its relation with nucleon transfer. The latter is usually quantified in asymmetric collisions from average numbers of neutrons and protons in each reaction product. Methods to extract probabilities to find a given number of protons and neutrons in the fragments were also developed to better evaluate the predictive power of TDHF simulations \cite{koonin1977,simenel2010}. However, the resulting mass and charge distributions were found to be too narrow in deep-inelastic collisions \cite{dasso1979} due to a lack of quantum fluctuations. The latter are included in beyond mean-field approaches such as time-dependent random phase approximation (TDRPA) \cite{balian1984} and stochastic mean-field (SMF) methods \cite{ayik2008,lacroix2014,ayik2025SMFreview}. TDDFT is now one of the main tools to predict formation of heavy and super-heavy neutron rich nuclei in deep-inelastic collisions. % (see \cite{nakatsukasa2016,simenel2018,sekizawa2019,godbey2020,simenel2025a}. 
\begin{marginnote}
\entry{TKE}{Total kinetic energy of outgoing fragments at large distances.}
\entry{TKEL}{Total kinetic energy loss refers to the difference between initial kinetic energy and TKE.}
\entry{SMF}{Stochastic mean-field approaches include quantum fluctuations.}
\entry{TDRPA}{The time-dependent random-phase approximation includes quantum fluctuations assuming they are small.}
\end{marginnote}

\subsubsection{Dissipation from transfer mechanisms}

Multi-nucleon transfer plays a crucial role in reactions between heavy ions such as those used to form SHE. 
In particular, it is expected to be the main mechanism for dissipation between the reactants \cite{randrup1978}.
The least bound nucleons, i.e., those closest to the Fermi surface, are easier to transfer than those which are more bound, thus contributing more rapidly to dissipation.
In particular, in neutron-to-proton asymmetric collisions, one expects a rapid flow of protons and neutrons in opposite ways to equilibrate this asymmetry \cite{simenel2001,umar2017}. Such equilibration occurs within a timescale of about $10^{-21}$~s \cite{jedele2017,simenel2020}, which is of the same order as the timescale for dissipation obtained from TKEL in TDHF simulations \cite{simenel2020}, thus confirming that transfer and dissipation are likely to impact each other. As a result, such isospin dynamics have been shown to impact fusion barriers \cite{godbey2017}.

Reactions with stable doubly-magic nuclei, which are well bound and thus less able to transfer their nucleons, should then lead to less dissipation and thus be able to reach more compact shapes in the PES entry point. Note, however, that the effect of magicity is expected to rapidly disappear in neutron-to-proton asymmetric reactions due to the rapid isospin equilibration process \cite{simenel2012b,mohanto2018}. 
This could explain the success of $^{48}$Ca induced reactions on actinides to form SHE via hot fusion reactions (cf.\ Sec.~\ref{sec:cold_and_hot_fusion}). Indeed, $^{48}$Ca is not only doubly magic, its neutron-to-proton ratio is also similar to the actinide targets. Experiments to form $Z=120$ with $^{50}$Ti beams are then expected to be more challenging as these conditions are not as well fulfilled with this single magic and less neutron rich beam.

\subsubsection{transfer probabilities with particle number projection technique in TDHF}

Since the TDHF wavefunction after collision is not an eigenstate of the particle-number operator for one of the reaction products, it is a superposition of states with different particle numbers in the fragments. One can extract the probability of finding $n$ nucleons in a spatial volume $V$ using the particle-number projection operator,
\begin{equation}
\hat{P}_n^{(q)} = \frac{1}{2\pi}\int_0^{2\pi}e^{i(n-\hat{N}_V^{(q)})}\dd\theta,
\end{equation}
where $\hat{N}_V^{(q)}$ is the number operator in the spatial region $V$ that contains one of the reaction products. This method has been routinely used to project out a good particle-number state from HFB or BCS states in nuclear structure calculations.

The particle-number projection method was first applied for head-on collisions of $^{16}$O+$^{208}$Pb \cite{simenel2010}, showing that one proton transfer probability at subbarrier energies is in good agreement with experimental data. Later, in Ref.~\cite{sekizawa2013}, the particle-number projection method was used to extract transfer probabilities as a function of the impact parameter. Production cross sections of primary fragments are directly calculated as
\begin{equation}
\sigma_{N,Z} = 2\pi\int_{b_\text{min}}^{b_\text{max}}P_{N,Z}(b)\,\dd b,
\end{equation}
where $P_{N,Z}(b)=P_N^{(n)}(b)P_Z^{(p)}(b)$ is the probability to form a reaction product with $N$ neutrons and $Z$ protons. $b_\text{min}$ denotes the minimum impact parameter for binary reactions, inside which fusion reactions take place, while $b_\text{cut}$ denotes a cutoff impact parameter at which transfer probabilities are negligible. The method was applied for $^{40,48}$Ca+$^{124}$Sn, $^{40}$Ca+$^{208}$Pb,  $^{58}$Ni+$^{208}$Pb~\cite{sekizawa2013}, and $^{58}$Ni$+^{124}$Sn \cite{wu2019} systems, showing that nucleon transfer cross sections for a few nucleon transfers are quantitatively described within TDHF. It was also used to predict production cross-sections in the $^{100}$Sn region~\cite{wu2022}. The particle-number projection technique is now a standard tool to study fragment mass and charge distributions in heavy-ion collisions (see \cite{simenel2025a} and references therein).

\subsubsection{Quantum fluctuations in deep-inelastic collisions}

However, the agreement becomes less accurate as the number of transferred nucleons increases far apart form the average value. It is due to the well-known limitation inherent to the mean-field description, where fluctuations of one-body observables are  severely underestimated. It can be shown, based on an extended variational principle of Balian and V\'en\'eroni, that the variational space of TDHF is optimized for expectation values of one-body operators, while the variance $\bigl<\hat{\mathcal{O}}^2\bigr>-\bigl<\hat{\mathcal{O}}\bigr>^2$, which contains an operator of two-body character, is out of the variational space used to derive TDHF \cite{balian1981}. To better describe processes far from the average values, one needs to employ extended approaches such as TDRPA \cite{balian1984} or SMF theory \cite{ayik2008,lacroix2014,ayik2025SMFreview} to take into account effects of one-body fluctuations and correlations. Both methods have been successfully applied for various reaction systems (see, e.g., \cite{simenel2011,williams2018,gao2025,sekizawa2020,godbey2020b}), often showing quantitative agreements with experimental data. 
%It indicates one-body dissipation, fluctuations and correlations are predominant mechanisms in low-energy heavy-ion reactions.

\subsubsection{Multi-nucleon transfer in actinide collisions}

Actinide collisions have been proposed as a possible way to form neutron-rich heavy and superheavy nuclei via multi-nucleon transfer, making use of their large neutron-to-proton ratio. One could potentially explore the expected neutron shell closure $N=184$ with this technique. 

However, TDHF calculations of such systems are numerically challenging due to the large number of nucleons and the fact that both collision partners are usually prolately deformed. Indeed, evaluations of observables then  require a double integral on the orientation of their deformation axis to account for all possible initial orientations. For these reasons, actinide collisions have essentially been studied for selected orientations  in order to investigate the reaction mechanisms and find out which orientations (if any) could lead to the formation of heavy and superheavy nuclei \cite{golabek2009,kedziora2010,umar2018,ayik2020,zhang2024}.

\begin{figure}
\includegraphics[width=15cm]{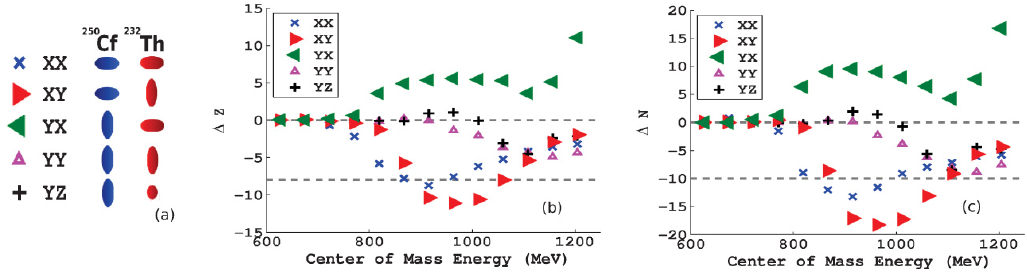}
\caption{ TDHF studies of $^{232}$Th$+^{250}$Cf central collisions for selected orientations represented in (a). 
Change in (b) proton and (c) neutron numbers in the $^{250}$Cf-like fragment are shown as a function of center-of-mass energy. $\Delta Z,N>0$ indicates an inverse quasi-fission, essentially observed for the $YX$ orientation. Adapted from \cite{kedziora2010}.}
\label{fig:ThCf} 
\end{figure}

The findings of these studies can be summarized as follows. Contact times rarely exceed 4~zs and are highest for collisions between the sides of the prolately deformed actinides. The largest average transfer of nucleons (i.e., obtained from expectation values of particle numbers in the fragments, not accounting for quantum fluctuations) are obtained from tip-on-side collisions. For such orientations, a net transfer of nucleons is observed from the tip to the side. 

An example of orientation and energy dependent multi-nucleon transfer is shown in Fig.~\ref{fig:ThCf} for  $^{232}$Th$+^{250}$Cf central collisions at selected orientations \cite{kedziora2010}.
A net transfer from the heavy to the light fragment is observed at $E_{c.m.}\gtrsim800$~MeV except for the $YX$ orientation corresponding to a collision of the tip of $^{232}$Th with the side of $^{250}$Cf. In this latter case, a net transfer is observed towards the heavy fragment, a process also called ``inverse quasi-fission'' (IQ). 
An interesting feature of such reactions is that the amount of transfer varies relatively slowly with energy. 
Nevertheless, it is desirable to aim at the lowest possible energy to optimize the survival probability of the heavy fragment against fission, and to minimize the number of evaporated neutrons so that the most neutron rich heavy nucleus could be produced. Indeed, the results shown in Fig.~\ref{fig:ThCf} correspond to primary fragments prior to statistical decay via fission and evaporation. 

Making use of increase in computational power, further TDHF studies should now be performed for such systems accounting for more intermediate orientations and varying impact parameters in order to evaluate the cross-sections for primary fragment productions. Statistical decay of these fragments should also be included to compute their survival probabilities. Such studies are necessary to evaluate the ability of actinide collisions to compete with fusion-evaporation for the formation of neutron rich heavy and superheavy nuclei.

\subsection{Quasi-fission}

Quasi-fission refers to the formation of fission-like fragments following mass transfer and without formation of an equilibrated compound nucleus \cite{toke1985}.
Optimization of the entrance channel reaction to form SHE requires a good understanding of quasi-fission, motivating numerous studies (see \cite{hinde2021} and \cite{simenel2018} for reviews on experimental and theoretical aspects, respectively).  %\cite{wakhle2014,oberacker2014,ayik2015a,umar2015a,hammerton2015,sekizawa2016,umar2016,morjean2017,yu2017,guo2018c,zheng2018,godbey2019,simenel2021,li2022,stevenson2022,mcglynn2023,lee2024,scamps2024b,li2024c}.
Quasi-fission reactions are characterised by the full damping of the initial kinetic energy of the collision partners, a significant mass transfer at contact, and contact times of the order of few zeptoseconds ($1$~zs$=10^{-21}$~s) or few 10~zs, inducing significant correlations between mass and scattering angles of the fragments.

\subsubsection{Energy dissipation}

Typical timescales for quasi-fission are much larger than those of dissipation (which occur within $\sim1$~zs) \cite{simenel2020}. It is then expected that outgoing fragments are emitted with a kinetic energy roughly equal to that of their Coulomb repulsion energy at scission. In other words, quasi-fission fragments have similar total kinetic energy (TKE) as fission fragments as given by the Viola systematics. 

\begin{figure}
\resizebox{0.9\columnwidth}{!}{\includegraphics{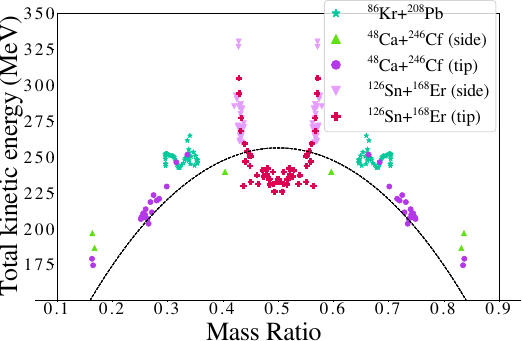}}
\caption{Total kinetic energy as a function of fragment mass ratio $A_1A_2/(A_1+A_2)$ for various reactions producing $^{294}$Og as a compound nucleus. The Viola systematics \cite{hinde1987} is shown with black dashed line. ``side'' and ``tip'' refer to orientations leading to collisions with the side and tip of the prolately deformed heavy target, respectively. From \cite{mcglynn2023}.}
\label{fig:TKE} 
\end{figure}

Figure \ref{fig:TKE} shows fragment total kinetic energies calculated with TDHF for reactions forming $^{294}$Og as a compound nucleus \cite{mcglynn2023}. The outgoing quasi-fission fragments have a TKE that approximately follows Viola systematics \cite{hinde1987}, showing that the reactions are indeed fully dissipated. 

Nevertheless, deviations from Viola systematics are observed, with large TKE found for Sn fragments and, to a lesser extent, for Pb fragments. As in fission, this is interpreted as an influence of spherical shell effects which leads to more  compact systems at scission and thus a larger Coulomb repulsion being converted into TKE. An effect of the initial orientation of the heavy collision partner  on the TKE has also been observed in TDHF calculations of $^{48}$Ca$+^{249}$Bk, with collisions with the tip (respectively side) of $^{249}$Bk leading to a TKE slightly larger (smaller) than predicted by the Viola systematics \cite{umar2016}.

\subsubsection{Mass-equilibration}

Mass equilibration timescale being much larger than those of dissipation (which occur within $\sim1$~zs) \cite{simenel2020}, quasi-fission is not in itself a dissipation mechanism as the initial kinetic energy is already essentially fully dissipated when the mass equilibration kicks off. 
Nevertheless, quasi-fission remains the main mechanism competing with fusion, thus preventing SHE formation. 
A possible explanation is that the mass transfer essentially goes from the heavy to the light fragment (except in  cases of inverse quasi-fission induced by orientation and shell effects \cite{zagrebaev2006,kedziora2010,lee2024b}), producing a dinuclear system that is more symmetric than the entrance channel and thus increasing the Coulomb repulsion as the charge product increases. 
The resulting dynamical increase of the fusion barrier due to mass equilibration thus prevents compound nucleus formation and leads to the formation of fission like outgoing fragments. 

\begin{figure}
    \centering
    \includegraphics[width=0.9\linewidth]{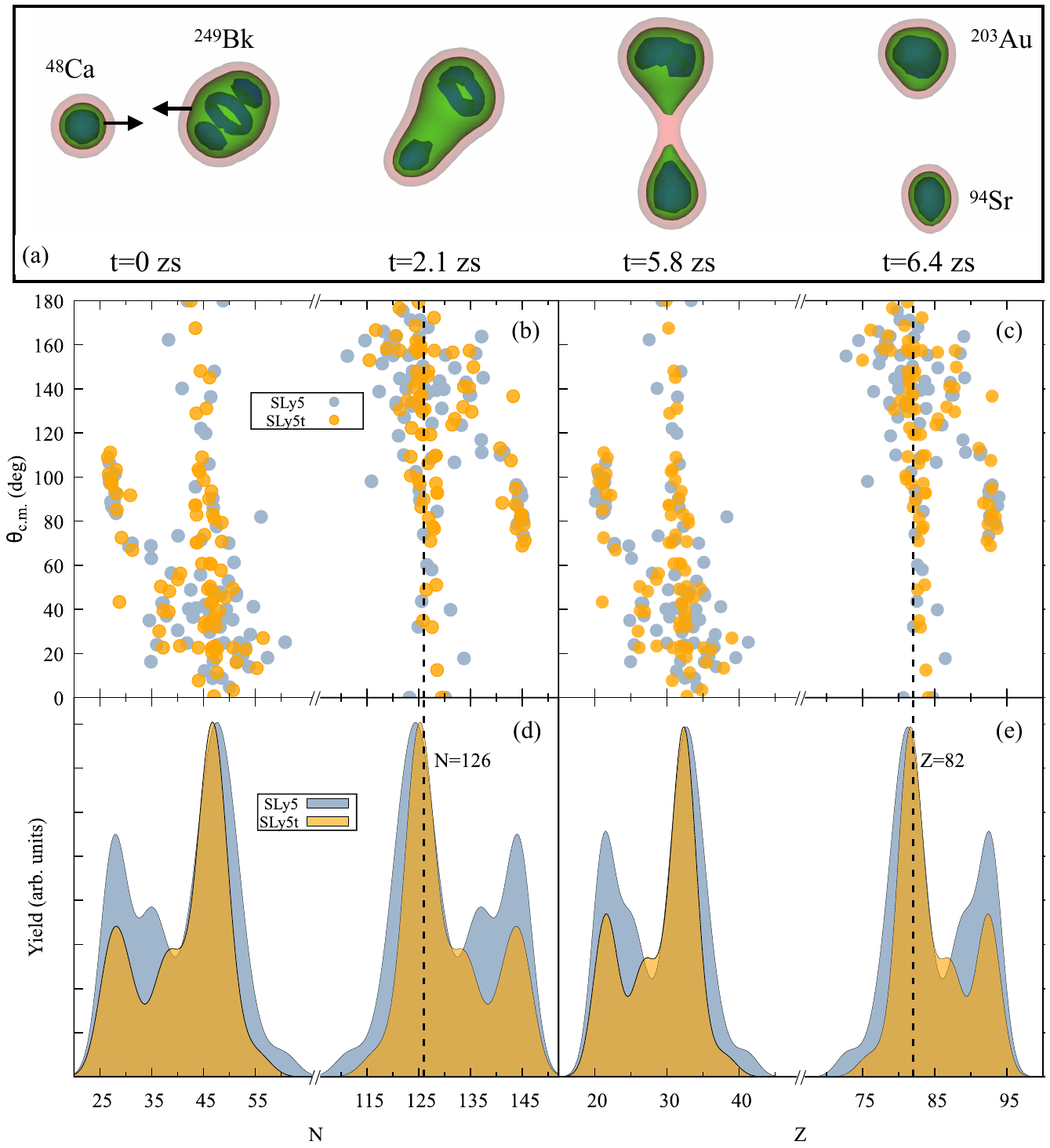}
    \caption{(a) Isodensity surfaces from a TDHF simulation of $^{48}$Ca$+^{249}$Bk at $E_{c.m.}=234$~MeV resulting into a quasi-fission reaction \cite{godbey2019}. (b) Neutron number-angle distributions (NAD) for two Skyrme EDF with (SLy5) and without (SLy5t) tensor force. (c) Same for proton numbers (ZAD). (d) and (e) show the projections on the fragment neutron $N$ and proton $Z$ numbers, respectively. (b-e) are from Ref.~ \cite{li2022}.}
    \label{fig:CaBk}
\end{figure}

An example of TDHF simulation of quasi-fission reaction in $^{48}$Ca$+^{249}$Bk is shown in Fig.~\ref{fig:CaBk}(a) \cite{godbey2019}. The exit channel is less mass asymmetric than the entrance channel. However, the  mass equilibration is partial as the system separates after $\sim5$~zs while, according to TDHF systematics, the time for full equilibration is expected to be of the order of $\sim20$~zs \cite{simenel2020}. 

TDHF predictions of fragment neutron and proton number distributions for this reaction are shown in Fig.~\ref{fig:CaBk}(d,e), respectively \cite{li2024c}. 
The most asymmetric peaks correspond to deep inelastic collisions (quasi-elastic events are excluded) with no or little average mass transfer between the reactants. Less asymmetric events are associated with quasi-fission through a partial mass equilibration via multi-nucleon transfer. 
Large peaks are observed near $^{208}$Pb and its complementary light fragment. 
Theoretically, these peaks are usually associated with $Z=82$ and $N=126$ shell effects in the heavy fragment \cite{wakhle2014,hinde2018,morjean2017}, though the $N=56$ octupole deformed shell effect in the light fragment has also been invoked \cite{godbey2019}. Note that, although the tensor force can affect the competition between these shell effects \cite{li2022,li2024c}, they are likely to contribute to the same theoretical mode as defined by the presence of a fission valley in the PES \cite{mcglynn2025}. Similar shell effects have also been predicted in reactions forming actinides \cite{simenel2021,lee2024b,simenel2025c}, where the same octupole deformed shell effects as in fission \cite{scamps2018,scamps2019} are expected to be at play.
The standard interpretation is that the mass symmetrization process gets stopped when the system falls into a fission valley induced by these shell effects. 

Experimentally, the role of shell effects in quasi-fission is still debated. 
While {\it ``clear evidence of quantum shell effects in slow quasi-fission processes,''} was reported in ~\citep{pal2024} for fragments near $A \approx 96$, it has been argued that the observed $A \simeq 208$ peak might result from sequential fission of the target-like fragment~\citep{jeung2022}. Similarly, in reactions leading to actinide compound nuclei (rather than SHE), {\it ``clear evidence''} of shell-stabilized fragments has been reported in~\citep{chizhov2003}, while others find only {\it ``weak evidence''}~\citep{hinde2022}.

It is possible that the lack of two-body mechanisms in TDHF artificially enhances the influence of shell effects in the dynamics. 
Though computationally demanding, beyond mean-field calculations are required to confirm or infirm TDHF predictions of an important role of shell effects in quasi-fission. 
Although a detailed understanding of quasi-fission dynamics is not required to better predict SHE formation (whether or not shell effects stop mass equilibration should not impact the probability of forming a compound nucleus), it is important to test our ability to predict outcomes of complex dynamical processes such as fission and quasi-fission in order to improve theoretical models that can then be used to select optimum entrance channels for SHN formation.

\subsubsection{Mass-angle correlations}

Mass and charge distributions alone are not sufficient to infer timescales on which quasi-fission takes place. 
Mass-angle distributions (MAD), however, exhibit correlations that can be used to evaluate contact times between the fragments \cite{durietz2011,durietz2013}. MADs have been useful to quantify non-equilibrium processes from slower processes such as fusion-fission, and thus provide a tool to evaluate compound nucleus formation probability \cite{banerjee2019}.
They have also been used to identify the entrance channels with the longest contact times to optimize the formation of $Z=120$ SHE \cite{albers2020}, and to separate fission and quasi-fission modes in superheavy systems \cite{banerjee2021}.
\begin{marginnote}
\entry{MAD}{Mass angle distributions show correlations between  scattering angle and  nucleon number, reflecting sticking time and shape evolution of the reactants.}
\end{marginnote}

Mass-angle correlations are present because the timescale for mass equilibration is of the order of typical timescale for the di-nuclear system to undergo a full rotation. Correlations are then present for contact times of the order of few zs, though they are  expected to wash out in slow quasi-fission (typically more than 20~zs) where the system often undergoes more than one full rotation. 

Of course, extracting the rotation velocity from experimental data is a difficult task as it depends on the angular momentum which is unknown for individual collisions, though average angular momentum could be extracted from MAD subtraction at different energies \cite{tanaka2021}. TDHF simulations can then be used to pinpoint the expected angular momentum associated with specific positions in the MAD \cite{wakhle2014,hammerton2015,guo2018c}. These simulations also help interpret orientation effects in the MAD, i.e., which regions of the MAD are populated by collisions with the side or tip of a deformed actinide target. TDHF calculations have also been used to evaluate the moment of inertia of the rotating di-nuclear system in order to extract contact times phenomenologically \cite{prasad2016,prasad2017}.

Figures \ref{fig:CaBk}(b,c) show neutron (proton) number-angle distributions from TDHF simulations of $^{48}$Ca$+^{249}$Bk at near barrier energy, respectively called NAD and ZAD \cite{li2024c}. Each point corresponds to a single TDHF calculation. Thus, such investigations are very computationally demanding. We clearly observe an accumulation of events in the bottom left and top right of the NAD and ZAD, confirming the presence of strong correlations between fragment composition and scattering angles. The calculations also predict a stop of the drift towards mass symmetry, interpreted in terms of shell effects stopping mass equilibration.

\section{Formation and decay of the compound nucleus \label{sec:CN}}

The formation of a compound nucleus requires avoiding quasi-fission.
TDHF calculations can help identify which orientations and energies lead to a full capture of the collision partners and the formation of a compact compound system. 
The compound nucleus then needs to survive against fission in order for a cold SHN to be formed. 
Here, we study the formation and decay of superheavy compound nuclei. We first investigate TDHF trajectories on the PES to identify what conditions can lead to the formation of a compact system. 
We then investigate the cases of cold and hot fusion reactions  with TDHF calculations combined with statistical decay modeling. 
Finally, fission modes of the SHN are briefly discussed.

\subsection{Evolution of multipole moments on PES}

The computation of multipole moments such as the quadrupole $Q_{20}$ and octupole $Q_{30}$ moments, usually used to quantify elongation and asymmetry, respectively, are standard in TDHF. For non-central collisions, they require the determination of the principal axis from the diagonalisation of the quadrupole tensor $Q_{ij}$, with $i,j=x, y$ or $z$. 
Once the evolutions of $Q_{20}(t)$ and $Q_{30}(t)$ have been calculated, it is straightforward to overlay them on a previously computed PES to search for eventual correlations between the TDHF trajectories in the $Q_{20}-Q_{30}$ plane and the PES topography (in particular its fission valleys). 

This was recently done  for reactions forming $^{226}$Th \cite{lee2024b}. 
The TDHF trajectories were found to be influenced by the presence of fission valleys. 
Surprisingly, the $^{96}$Zr$+^{130}$Sn, which is more symmetric than the asymmetric fission mode of  $^{226}$Th, was found to encounter an inverse quasi-fission which could be interpreted as an influence of the asymmetric fission valley.

A similar study  was also done for $^{48}$Ca$+^{246}$Cf, $^{86}$Kr$+^{208}$Pb and $^{126}$Sn$+^{168}$Er central collisions forming $^{294}$Og as a compound nucleus \cite{mcglynn2023}. 
No compound nucleus formation was found in $^{86}$Kr$+^{208}$Pb, in $^{126}$Sn$+^{168}$Er, and in tip orientations of $^{48}$Ca$+^{246}$Cf, even at energies as high as $60-80\%$ above the barrier. TDHF trajectories in the $Q_{20}-Q_{30}$ plane showed that, indeed, the entry point is always found outside the conditional saddle point, with the elongation of the system further increasing towards quasi-fission as time increases. 
Only side orientations of $^{48}$Ca$+^{246}$Cf were found to fuse. 
This can be seen in Fig.~\ref{fig:294Og} where entry points (represented by stars) that are located at smaller elongations than the conditional saddle point ($Q_{20}^{sad.}\simeq33$~b) produce trajectories that get trapped and can be associated with fusion.

In reactions forming SHN, it is then desirable to reach the smallest possible elongation at contact which is usually achieved with actinide targets that are prolately deformed. In TDHF, collisions with the side of the target are usually the only ones that lead to fusion \cite{wakhle2014,oberacker2014,umar2016,guo2018c,godbey2019,mcglynn2023}. 
This is an important information to account for in choosing beam energies as collisions with the side of the actinide target are associated with a higher potential energy than the average Coulomb barrier. One could be tempted to select collisions with the tip which have a smaller Coulomb repulsion, and thus would form a compound nucleus at lower excitation energy with better chance to survive against fission. According to TDHF simulations, however, this would be a poor choice as tip orientations inevitably lead to quasi-fission.

\subsection{Cold and hot fusion reactions}\label{sec:cold_and_hot_fusion}

Reactions with a light projectile (e.g., $^{48}$Ca or $^{50}$Ti) on actinides are called ``hot fusion reactions'' while reactions on $^{208}$Pb are ``cold fusion''  reactions. The naming essentially refers to the difference in excitation energy of the compound nuclei in both types of reactions. 
\begin{marginnote}
\entry{Cold fusion}{Fusion reactions forming heavy nuclei at low excitation energy by emitting few neutrons, minimizing compound nucleus heating.}
\entry{Hot fusion}{Fusion reactions forming heavy nuclei at high excitation energy, emitting multiple neutrons during compound nucleus de-excitation.}
\end{marginnote}

\begin{figure}
    \centering
    \includegraphics[width=0.9\linewidth]{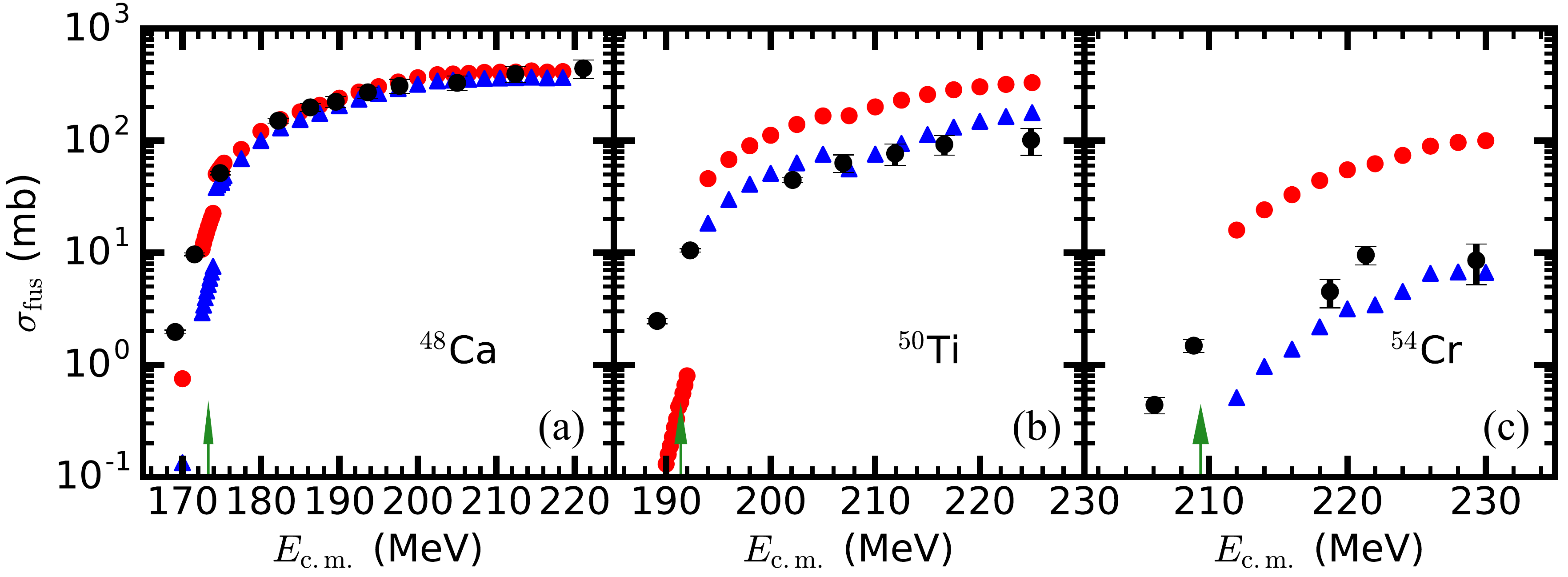}
    \caption{Fusion cross sections for (a) $^{48}$Ca$+^{208}$Pb, (b) $^{50}$Ti$+^{208}$Pb and (c) $^{54}$Cr$+^{208}$Pb. Experimental data (black circles) are from \cite{banerjee2019}.
    Red circles and blue triangles represent theoretical results from the combined TDHF and FdB approach, and obtained with different choices of injection points (see text). The arrows show the position of the capture thresholds.  Adapted from \cite{sun2022b}.}
    \label{fig:cold}
\end{figure}

Cold fusion reactions $^{48}$Ca,$^{50}$Ti and $^{54}$Cr$+^{208}$Pb have been studied in \cite{sun2022b}, with the emphasis in describing the fusion cross-section 
$$\sigma_\text{fus}(E)=\frac{\pi}{k^2}\sum_J(2J+1)T_J(E)P_\text{CN}(E,J).$$
Here, $T_J(E)$ is the transmission probability at centre of mass energy $E$ and angular momentum $J$. It is computed using the coupled channel code \textsc{ccfull} \cite{hagino1999}. 
The compound nucleus formation probability is then evaluated within the FbD model \cite{cap2011}.
The resulting fusion cross-sections are reproduced in Fig.~\ref{fig:cold}. 
A good agreement with experimental data is found, though the latter somewhat depends on the choice made for the injection (or entry) point obtained from TDHF \cite{sekizawa2019b} and which defines the initial condition for the FbD calculation. 
Specifically, the red circles in Fig.~\ref{fig:cold} are obtained with an injection point defined using empirical radii $R=r_0A^{1/3}$ while the blue triangles are obtained using Hartree-Fock radii (see \cite{sun2022b} for details). A better agreement with experimental data is obtained with the latter choice, in particular for the heavier systems. 

Hot fusion has also been studied with this method \cite{sun2023b}. 
An example of TDHF study of hot fusion reactions can also be found in \cite{guo2018c} for the $^{48}$Ca$+^{239,244}$Pu reactions. 
Here, the aim was to explain the experimental observation that fusion-evaporation cross sections in the $3n$ reaction channel 
are 50 times lower with $^{239}$Pu than with $^{244}$Pu \cite{utyonkov2015}. 
A combination of TDHF simulations with statistical model code HIVAP calculations of the survival probabilities showed that, indeed, quasi-fission is reduced with $^{244}$Pu and the survival probability is enhanced due to a larger fission barrier in the more neutron-rich compound nucleus. 

These studies demonstrate that combining TDHF with other methods such as coupled-channels, transport models, and statistical decay codes provide a powerful approach to obtain quantities that can be directly compared with experimental data. 
Nevertheless, each approach has its own limitations and sensitivity to initial conditions, as demonstrated by the choice of the injection point for the FbD model. The predictive power of the combined approach is therefore limited by phenomenology. It is then desirable to pursue efforts towards beyond mean-field dynamical models that would allow for a complete description of the formation of the compound nucleus, and eventually its subsequent decay.

\subsection{Fission modes of SHN}
\label{sec:fission}

SHN provide an interesting ground for fission studies. 
While fission properties of actinides and sublead nuclei have been thoroughly studied, those of SHN are only predicted by theory and still require experimental investigations. A deeper understanding of fission in this region will help constrain theoretical models and improve their predictive power in regions that are experimentally inaccessible, such as heavy neutron-rich nuclei at the end of the $r$-process that are expected to reach a region of spontaneous fission \cite{goriely2015b}.

The PES shown in Fig.~\ref{fig:294Og} for  $^{294}$Og \cite{mcglynn2023} exhibits a very asymmetric  valley, associated with a super-asymmetric fission mode.
Indeed, the  adiabatic asymmetric fission path (dashed line) follows closely the bottom of this valley. 
It begins just before the second barrier (around $Q_{20}\simeq35$~b) and extends up to scission.
The heavy fragment formed at scission is close to the doubly-magic $^{208}$Pb nucleus.
Although a succession of proton and neutron shell effects could be at play as in the actinide region \cite{bernard2023}, this indicates that the final asymmetry is influenced by $Z=82$ and $N=126$ spherical shell gaps. 
We also see that the scission configuration in the asymmetric mode occurs at relatively small values of $Q_{20}$. 
This is compatible with the expected influcence of shell effects in the heavy fragment leading to a compact scission configuration, and should then be associated with large TKE.
Such superasymmetric fission mode influenced by $^{208}$Pb shell effects has been predicted in various theoretical works \cite{warda2018,matheson2019,ishizuka2020}.

Finally, the strong similarity predicted between fission and quasi-fission modes in actinides \cite{simenel2021,lee2024b,simenel2025c} could open ways to investigate at least some properties of SHN fission by studying their quasi-fission modes \cite{mcglynn2023}. These properties could be, e.g., average mass and charge of the  fragments in the superasymmetric mode, as well as their TKE. If more than one mode is present, however, then the competition between these modes could not be inferred from quasi-fission studies \cite{lee2024b} and proper fission of the compound nucleus would have to be measured, which is a considerable challenge in view of the small fusion cross-sections and the difficulty to separate fission from quasi-fission.

\section{Conclusion and perspectives \label{sec:conclusion}}

The theoretical exploration of SHN continues to push the limits of nuclear many-body physics.
Significant progress has been achieved through TDDFT, offering a microscopic description of reaction dynamics from the initial collision to compound nucleus formation.
By combining TDHF-based simulations with complementary approaches—such as coupled-channel models, Langevin dynamics, and statistical decay codes—quantitative links with experimental data have become increasingly robust.
These hybrid frameworks have clarified the mechanisms of fusion hindrance, quasi-fission, and multi-nucleon transfer, deepening our understanding of the competing pathways that govern SHN synthesis.

Despite these advances, major theoretical challenges remain.
Current mean-field approaches omit two-body correlations and fluctuations essential to describe compound nucleus formation.
Beyond-mean-field methods incorporating stochastic dynamics and quantum fluctuations are thus a critical next step toward a predictive description of heavy-ion reactions.
Further developments in fluctuation–dissipation extensions of TDHF will help quantify mass, charge, and energy diffusion during quasi-fission and transfer processes.

The fission properties of the heaviest systems remain uncertain.
Microscopic calculations highlight the role of shell effects in shaping scission configurations, asymmetric valleys, and the connection between fission and quasi-fission modes.
Experimental confirmation—especially of superasymmetric fission channels linked to the doubly magic $^{208}$Pb fragment—will be decisive for validating and refining current models.

Looking ahead, the synthesis of new elements and neutron-rich superheavy isotopes will benefit from tighter integration of experiment and theory.
Advances in beam and detector technologies, coupled with  microscopic simulations, will guide the choice of target–projectile combinations and beam energies—not only for traditional fusion–evaporation reactions, but also for actinide collisions forming heavy neutron-rich nuclei.

\begin{summary}[SUMMARY POINTS]
\begin{enumerate}
  \item Fusion--evaporation reactions remain the only experimentally successful method to produce superheavy nuclei, but their cross sections are extremely small because compound nucleus formation is strongly hindered by quasi-fission and survival against fission is rare.

  \item Time-dependent density functional theory (TDDFT), in particular time-dependent Hartree--Fock (TDHF) and its extensions, provides a microscopic framework to describe the early stages of heavy-ion collisions leading to capture, fusion, transfer, or quasi-fission, with modern three-dimensional implementations including tensor forces and, in some cases, dynamical pairing.

  \item Density-constrained methods (DC-TDHF, DCFHF, and FHF) allow the extraction of microscopic nucleus--nucleus interaction potentials and clarify the respective roles of dynamical effects and Pauli repulsion in heavy systems.

  \item Fusion hindrance in heavy and superheavy systems manifests as an ``extra-push'' energy above the static fusion barrier; microscopic TDHF calculations reproduce the onset of this effect and predict smaller extra-push energies than macroscopic models.

  \item Quasi-fission is the dominant process competing with fusion in reactions forming superheavy nuclei and is characterized by strong energy dissipation, short contact times, and incomplete mass equilibration, with a strong dependence on nuclear deformation, orientation, and shell effects.

  \item Multi-nucleon transfer reactions offer an alternative pathway to produce neutron-rich heavy and superheavy nuclei, potentially accessing regions near the predicted $N = 184$ shell closure.

  \item Standard TDHF underestimates fluctuations in fragment mass and charge distributions; beyond-mean-field approaches such as the time-dependent random-phase approximation and stochastic mean-field methods improve the description of multi-nucleon transfer and deep-inelastic collisions.

  \item Because TDHF does not describe quantum tunneling, full thermalization, or statistical decay, hybrid approaches coupling microscopic dynamics to coupled-channels calculations, Langevin dynamics, fusion-by-diffusion models, and statistical decay codes are used to connect theoretical predictions with experimental observables.
\end{enumerate}
\end{summary}

\begin{issues}[FUTURE ISSUES]
\begin{enumerate}
  \item Fully microscopic descriptions of superheavy element formation will require systematic inclusion of beyond-mean-field effects, in particular quantum fluctuations and correlations that govern mass, charge, and angular-momentum distributions.

  \item Dynamical pairing correlations need to be consistently incorporated into time-dependent simulations of fusion, quasi-fission, multi-nucleon transfer, and fission to assess their impact on reaction outcomes across the superheavy region.

  \item A unified treatment of entrance-channel dynamics, compound nucleus formation, and decay—using consistent energy density functionals and model assumptions—remains essential for reducing uncertainties in predicted production cross sections.

  \item Improved microscopic input for hybrid approaches, including injection points, friction coefficients, and diffusion parameters, is required to better constrain Langevin, fusion-by-diffusion, and statistical decay models.

  \item Microscopic predictions for multi-nucleon transfer reactions should be extended to include secondary decay processes in order to enable direct and quantitative comparisons with experimental fragment yields.
\end{enumerate}
\end{issues}

\section*{Acknowledgement}
This work has been supported by the Australian Research Council Discovery Project (project number DP190100256) funding schemes, National Natural Science Foundation of China (Grants No. 12435008, No. 12375127, and No. 12205308), the Strategic Priority Research Program of the Chinese Academy of Sciences (Grants No. XDB34010000 and No. XDB1550000), and JSPS Grant-in-Aid for Scientific Research (Grant No. JP25H01269).
\bibliographystyle{ar-style5.bst}
\bibliography{VU_bibtex_master.bib}

\end{document}